\documentclass[fleqn,twoside,11pt]{article}

\usepackage{amssymb}
\usepackage{graphicx}
\usepackage{amsmath}

\begin{document}
\noindent
{\sf University of Shizuoka}

\hspace*{13cm} {\large US-06-03R}

\vspace{3mm}

\begin{center}

{\Large\bf  Tribimaximal Neutrino Mixing and a Relation}\\[.1in]
{\Large\bf  Between Neutrino- and Charged  Lepton-Mass Spectra}\\[.1in]

\vspace{3mm}
{\bf Yoshio Koide}

{\it Department of Physics, University of Shizuoka, 
52-1 Yada, Shizuoka 422-8526, Japan\\
E-mail address: koide@u-shizuoka-ken.ac.jp}

\date{\today}
\end{center}

\begin{abstract}
Brannen has recently pointed out that the observed charged 
lepton masses satisfy the relation $m_e +m_\mu +m_\tau = 
\frac{2}{3} (\sqrt{m_e}+\sqrt{m_\mu}+\sqrt{m_\tau})^2$, 
while the observed neutrino masses satisfy the relation 
$m_{\nu 1} +m_{\nu 2} +m_{\nu 3} 
= \frac{2}{3} (-\sqrt{m_{\nu 1}}+\sqrt{m_{\nu 2}}
+\sqrt{m_{\nu 3}})^2$.
It is discussed what neutrino Yukawa interaction form is favorable 
if we take the fact pointed out by Brannen seriously.
\end{abstract}

\vspace{3mm}

{\large\bf 1 \ Introduction}

It is well-known that the observed charged lepton mass spectrum 
\cite{PDG04} satisfies the relation 
\cite{Koidemass,Koide90} 
$$
m_e+m_{\mu}+m_{\tau}=
\frac{2}{3}\left( \sqrt{m_e}+\sqrt{m_\mu}+\sqrt{m_{\tau}} 
\right)^2 , 
\eqno(1.1)
$$
with remarkable precision, while the observed 
masses $(m_{f1}, m_{f2}, m_{f3})$ of the other 
fundamental particles $f_i$
(quarks and neutrinos) do not satisfy \cite{othermass} a relation similar to
(1.1) straightforwardly.
However, Brannen \cite{Brannen} has recently pointed out a possibility that 
the observed neutrino masses can satisfy the relation 
$$
m_{\nu 1}+m_{\nu 2}+m_{\nu 3}=
\frac{2}{3}\left( -\sqrt{m_{\nu 1}}+\sqrt{m_{\nu 2}}+\sqrt{m_{\nu 3}} 
\right)^2 . 
\eqno(1.2)
$$
Of course, we cannot extract the values of the neutrino mass ratios
$m_{\nu 1}/m_{\nu 2}$ and $m_{\nu 2}/m_{\nu 3}$ from the
neutrino oscillation data $\Delta m^2_{solar}$ and
$\Delta m^2_{solar}$ unless we have more information on the neutrino
masses,
so that we cannot judge whether the observed neutrino masses 
satisfy the relation (1.2) or not.
Nevertheless, it is worthwhile to examine the Brannen's speculation 
seriously, because it seems to bring a new view into a lepton mass 
matrix model if his conjecture is correct.

The Brannen's conjecture (1.2) is somewhat unforeseen, because 
since we have known the existence of the lepton flavor mixing,
we expect that the neutrino mass matrix structure will be
quite different from that of the charged leptons, so that
the neutrino mass spectrum will also be different from that
of the charged leptons.
If the charged lepton masses satisfy the relation (1.1),
the neutrino masses will not be able to satisfy the relation
(1.2) even if we allow the replacement $\sqrt{m_{\nu i}}
\rightarrow  -\sqrt{m_{\nu i}}$.
In order to understand the formula (1.1), for example, 
a model \cite{Koide90,KF96,KT96}
with a seesaw-type
mass generation mechanism \cite{UnivSeesaw} has been proposed: 
$$
M_f=m_L M^{-1}_F m_R ,
\eqno(1.3)
$$
where $M_F$ are hypothetical heavy fermion mass matrices
and, for the charged lepton sector,  the structure
$M_E \propto {\rm diag}(1,1,1)$
is assumed.
(For neutrino sector, Eq.~(1.3) reads 
$M_\nu =m_L M_N^{-1} m_L^T$.)
Here, the mass matrices $m_L$ and $m_R$ are given by 
$m_R \propto m_L \propto {\rm diag}(v_1, v_2, v_3)$, 
where $v_i$ are 
vacuum expectation values (VEVs) of 3-family 
scalars $\phi_{Li}$ ($\phi_{Ri}$) and 
the VEVs $v_i$ satisfy the relation
$$
v_1^2 + v_2^2 + v_3^2 = \frac{2}{3}
 \left( v_1 + v_2 + v_3 \right) ^2.
\eqno(1.4)
$$
Of course, we do not rule out a possibility  that 
the values of $v_i$ are negative.
(Such a Higgs potential model which gives the relation
(1.4) is proposed in Ref.\cite{Koide90,KT96,Koide99,Koide06}.)
In such a seesaw-type model with diagonal $m_L$ and
$m_R$,  mixings among fermions $f_i$ are caused by 
the non-diagonal structure $M_F$.
If $M_E$ in the charged lepton sector is proportional
to a unit matrix {\bf 1}, the observed neutrino mixings
tell us that the structure of the heavy 
neutrino matrix (the Majorana mass matrix of the right-handed
neutrinos $N_R$) $M_N$ cannot 
be unit matrix, so that the eigenvalues of the matrix
$m_L M_N^{-1} m_L^T$ will not satisfy 
the relation (1.2) even if we allow the replacement
$\sqrt{m_{\nu i}} \rightarrow -\sqrt{m_{\nu i}}$.

If we take the Brannen's speculation seriously,
we must abandon such a model (1.3) (with a universal
$m_L$ structure).
For example, we may modify the model as
$$
M_f=m_L^f M^{-1}_F m^f_R ,
\eqno(1.5)
$$ 
where $M_F$  have a unit matrix structure universally,
at least, for the charged lepton and neutrino sectors,
and $m_L^f$ (and $m_R^f$) have flavor-dependent
structures.  
Then, in order to give the relations (1.1) and (1.2),
it is required that 
the eigenvalues $v_{fi}$ of the matrices $m^f_L$ ($m^f_R$) 
always satisfy the relation (1.4).
We relax the constraint on $m_L$ from a rigid ansatz of
the universal structure into the constraint (1.4).

Since $M_F \propto {\bf 1}$ do not play any essential role 
in the modified model (1.5), we may rather consider 
a Frogatt-Nielsen 
\cite{Frogatt} type model with six dimensional operators
$\overline{f}_{Li} \phi_{fi} H_L \phi_{fi} f_{Ri}$, 
where $H_L$ is the 
conventional SU(2)$_L$-doublet Higgs scalar, 
and $\phi_{fi}$ are 3-family SU(2)$_L$-singlet scalars. 
However, in the present paper, we do not have interest in
whether the effective mass matrix form originates in a seesaw 
model or in a Frogatt-Nielsen model.
Our interest is only in the neutrino Yukawa interaction form.
For convenience, in the present paper, we will
discuss a possible model in the framework of a seesaw 
mass matrix model.

The purpose of the present paper is not to give a model
which leads to the Brannen's conjecture.
The purpose is to investigate what Yukawa interaction form
is required when we accept the Brannen's conjecture and
when we take the observed neutrino mixing into consideration.
In the present paper, it is assumed that the eigenvalues 
$v_{fi}$ of the Dirac mass matrices $m_L^f$ of the charged 
leptons and neutrinos  satisfy the relation (1.4).
In the next section, we will introduce a useful parametrization 
for describing  charged lepton and neutrino masses in terms
of a permutation symmetry S$_3$ \cite{S3}.
In Sec.~3, we will investigate a possible neutrino mass
spectrum within the framework of the S$_3$ symmetry.
In Sec.~4, we will propose a phenomenological neutrino Yukawa
coupling form, where the S$_3$ symmetry is explicitly broken.
Finally, Sec.~5 is devoted in the concluding remarks.

\vspace{3mm}

{\large\bf 2 \ Mass spectrum parameters $z_i$ }

It is convenient to use parameters $z_i$
which are defined by $\langle \phi_{Li}^0 \rangle \equiv v_i
= v z_i$ with the normalization condition $z_1^2+z_2^2+z_3^2=1$.
The relation (1.1) [(1.4)] requires that the parameters $z_i$
satisfy the relation 
$$
z_1^2 + z_2^2 + z_3^2 = \frac{2}{3}
 \left( z_1 + z_2 + z_3 \right) ^2.
\eqno(2.1)
$$
The values of $z_i$ are obtained from the observed
charged lepton masses \cite{PDG04} as
$$
\frac{z_1}{\sqrt{m_e}} = \frac{z_2}{\sqrt{m_\mu}} =
\frac{z_3}{\sqrt{m_\tau}} =\frac{1}{\sqrt{m_e +m_\mu +m_\tau}},
\eqno(2.2)
$$
and the explicit numerical values are $z_1=0.016473$,
$z_2=0.236869$ and $z_3=0.971402$.

Since the relation (2.1) is invariant under any exchange
$z_i \leftrightarrow z_j$, it is useful to use the language
of the permutation symmetry S$_3$: We define the singlet
$\phi_\sigma$ and doublet $(\phi_\pi, \phi_\eta)$ of S$_3$ 
for the fields $\phi_i$ (and also for $f_i$):
$$
\phi_\pi =\frac{1}{\sqrt{2}} (\phi_3 -\phi_2), \ \
\phi_\eta =\frac{1}{\sqrt{6}} (2\phi_1 -\phi_2 -\phi_3), \ \
\phi_\sigma =\frac{1}{\sqrt{3}} (\phi_3 +\phi_2 +\phi_1), \ \
\eqno(2.3)
$$
where we have taken the basis $(e_1, e_2, e_3)$ as
$(e_1, e_2, e_3)=(e,\mu, \tau)$ corresponding to
the definition (2.2).
The definition (2.3) is different from the conventional one.
This definition will become useful for description of the
neutrino mixing as we show later [for example, see Eq.~(3.3)].

In order that the VEVs $v_i$ satisfy the relation (1.4), it
needs that the Higgs potential leads to the relation
$$
z_\pi^2 +z_\eta^2 = z_\sigma^2 ,
\eqno(2.4)
$$
where $\langle \phi_\pi^0\rangle = v z_\pi$, 
$\langle \phi_\eta^0\rangle = v z_\eta$ and 
$\langle \phi_\sigma^0\rangle = v z_\sigma$.
The explicit form of the Higgs potential which leads to 
the relation (2.4) has been given in Ref.\cite{Koide99,Koide06}.

By the way, in general, the eigenvalues $\lambda_i$ of any Hermitian matrix $M$ ($M M^\dagger$ if $M$ is not Hermitian) can be expressed by 
the following form:
$$
\begin{array}{l}
\lambda_1 = \frac{1}{3} a -  \frac{1}{3} b \sin\theta , \\
\lambda_2 = \frac{1}{3} a -  \frac{1}{3} b \sin\left(\theta +\frac{2}{3}\pi
\right) , \\
\lambda_3 = \frac{1}{3} a -  \frac{1}{3} b \sin\left(\theta +\frac{4}{3}\pi
\right) , \\
\end{array}
\eqno(2.5)
$$
where 
$$
a= {\rm Tr} M , \ \ \  b=\sqrt{2} \sqrt{ 3 {\rm Tr}(M^2)
-({\rm Tr}M)^2 } .
\eqno(2.6)
$$
Therefore, any three-flavor mass spectrum can always be
expressed by the two parameters $b/a$ and $\theta$ 
independently of the structure of $M$.
Only for the case with ${\rm Tr}(M^2) =\frac{2}{3}({\rm Tr}M)^2$,
it gives the relation $b=\sqrt{2} a$.
Inversely, if we take $b=\sqrt{2} a$ in (2.5), the matrix $M$
satisfies the relation ${\rm Tr}(M^2) =\frac{2}{3}({\rm Tr}M)^2$.
(Although the author \cite{Koide00} and Brannen \cite{Brannen}
have discussed  a specific mass matrix
form
$$
\frac{1}{\sqrt{3}} \left(
\begin{array}{ccc}
1 & 0 & 0 \\
0 & 1 & 0 \\
0 & 0 & 1
\end{array} \right) +
\frac{1}{\sqrt{6}} \left(
\begin{array}{ccc}
0 & e^{i\theta} & e^{-i\theta} \\
e^{-i\theta} & 0 & e^{i\theta} \\
e^{-i\theta} & e^{i\theta} & 0
\end{array} \right) ,
\eqno(2.7)
$$
in connection with the mass formula (1.1),
the constraint $b=\sqrt{2} a$ has been
assumed (not derived) in their models, 
so that  it does not mean that the relation (1.1) was 
derived in their models.)

In the present paper, since we take the three-flavor Higgs 
potential model \cite{Koide06} which gives the relation (2.1) 
[i.e. (2.4)], the parameters $z_i$ can be expressed as
$$
\begin{array}{l}
z_1 = \frac{1}{\sqrt{6}}  -  \frac{1}{\sqrt{3}}  \sin\theta_e , \\
z_2 = \frac{1}{\sqrt{6}} -  \frac{1}{\sqrt{3}}  
 \sin\left(\theta_e +\frac{2}{3}\pi \right) , \\
z_3 = \frac{1}{\sqrt{6}}  -  \frac{1}{\sqrt{3}}  
\sin\left(\theta_e +\frac{4}{3}\pi \right) , 
\end{array}
\eqno(2.8)
$$
where the numerical value of $\theta_e$ is given by
$$
\theta_e=\frac{\pi}{4} -\varepsilon =42.7324^\circ
\ \ \ (\varepsilon=2.2676^\circ),
\eqno(2.9)
$$
from the observed charged lepton masses.

Since we assume that the neutrino masses $m_{\nu i}$
satisfy the relation (1.2), 
we can also define the $z_i^\nu$ parameters in the neutrino 
sector as
$$
\begin{array}{l}
z_1^\nu = \frac{1}{\sqrt{6}}  -  \frac{1}{\sqrt{3}}  
\sin\theta_\nu , \\
z_2^\nu  = \frac{1}{\sqrt{6}} -  \frac{1}{\sqrt{3}}  
 \sin\left(\theta_\nu +\frac{2}{3}\pi \right) , \\
z_3^\nu  = \frac{1}{\sqrt{6}}  -  \frac{1}{\sqrt{3}}  
\sin\left(\theta_\nu +\frac{4}{3}\pi \right) , 
\end{array}
\eqno(2.10)
$$
with $|z_1^\nu|<|z_2^\nu|<|z_3^\nu|$.
Then, Brannen \cite{Brannen} has also empirically found 
that if the observed neutrino mass values are given by
$$
\theta_\nu =\theta_e +\frac{\pi}{12}  =57.7324^\circ ,
\eqno(2.11)
$$
which gives 
$$
z_1^\nu = -0.079938, \ \  z_2^\nu =0.385404, \ \ 
z_3^\nu =0.9192788. 
\eqno(2.12)
$$
The Brannen's empirical relation (2.11) [i.e. the
$z_i^\nu$-values (2.12)]  predicts
$$
R \equiv \frac{|m_{\nu 2}^2 -m_{\nu 1}^2|}{
|m_{\nu 3}^2 -m_{\nu 2}^2|} =
\frac{(z_2^\nu)^4 -(z_1^\nu)^4}{
(z_3^\nu)^4 -(z_2^\nu)^4} = 0.0318 ,
\eqno(2.13)
$$
which is in good agreement with the observed value of $R$
\cite{solar,atm}
$$
R=\frac{\Delta m_{solar}^2}{\Delta m_{atm}^2} =
\frac{(7.9^{+0.6}_{-0.5})\times 10^{-5}}{(2.72^{+0.38}_{-0.25})
\times 10^{-3}}
= (2.9\pm 0.5)\times 10^{-2} .
\eqno(2.14)
$$

However, note that we cannot predict the ratio $R$ only under
the assumption (1.2), i.e. unless we also assume the value of
$\theta_\nu$. 
In Fig.~1, we illustrate the predicted value of $R$ versus 
$\theta_\nu$ under the assumption (1.2).
The present observed values of $\Delta m^2_{ij}$ are
consistent with the value of $\theta_\nu \simeq 55^\circ
- 58^\circ$.
In other words, if we want to consider a model in which the
neutrino masses satisfy the relation (1.2), we must build
a model which gives $\theta_\nu \simeq 55^\circ - 58^\circ$.

In the present paper, we take the Brannen's conjecture (2.11) seriously,
and we investigate what Yukawa interaction from in the neutrino sector
can lead to the relation (2.11).
Brannen\cite{Brannen} has tried to build a new mass model with an
algebraic approach.
However, in the present paper, we will discuss a possible mass matrix 
form within the framework of a conventional mass matrix model,
i.e. based on a Higgs mechanism and an extended seesaw mechanism.

\vspace{3mm}

{\large\bf 3 \ S$_3$ symmetry and Yukawa interaction in the neutrino sector}

We have assumed that the Yukawa interaction in the charged lepton sector
is given by 
$$
H_e = y_e \left( \bar{\ell}_{L1} E_{R1} \phi^d_{L1} 
+\bar{\ell}_{L2} E_{R2} \phi^d_{L2} +
\bar{\ell}_{L3} E_{R3} \phi^d_{L3} \right),
\eqno(3.1)
$$ 
with a universal coupling constant $y_e$, where
$\ell_{Li} =(\nu_{Li}, e_{Li})$ and 
$\phi^d_{Li} =(\phi^0_{Li}, \phi^-_{Li})$. 
(In the previous section, the neutrino $\nu_i$ ($i=1,2,3$)
denoted the mass eigenstates.  
However, in the present section, we will use the same notation
$\nu_i$ as the SU(2)$_L$ partners of 
$e_{Li}=(e_{L}, \mu_{L}, \tau_{L})$, respectively.)
We also assume the same structure for 
$\bar{L}_{R i} e^c_{R i} \phi_{R i}$, where 
$L_{Ri} =(E^c_{Ri}, N^c_{Ri})$ and 
$\phi^d_{Ri} =(\phi^+_{Ri}, \bar{\phi}^0_{Ri})$.
The explicit quantum number assignments are found, for
example, in Refs.\cite{KT96,evl01} for SU(2)$_L \times$SU(2)$_R \times$U(1)
model, and in Ref.\cite{SO10} for SO(10)$\times$SO(10) model.

The Yukawa interaction (3.1) is invariant under the S$_3$
symmetry, but, of course, it is not a general form which
is invariant under the S$_3$ symmetry.
Only when we consider that the Yukawa interaction (3.1) with
the VEV $\langle \phi_{Li} \rangle = v_i= v z_i$ 
gives the mass matrix $m_L^e$ in the seesaw model (1.5) with 
$M = M_0 {\bf 1}$,
the predicted charged lepton masses satisfy the formula (1.1).

The present observed neutrino data strongly suggest that the neutrino mixing 
is almost described by the so-called tribimaximal mixing \cite{tribi}
$$
U=\left(\begin{array}{ccc}
\frac{2}{\sqrt6} & \frac{1}{\sqrt3} & 0 \\
-\frac{1}{\sqrt6} & \frac{1}{\sqrt3} & -\frac{1}{\sqrt2} \\
-\frac{1}{\sqrt6} & \frac{1}{\sqrt3} & \frac{1}{\sqrt2} \\
\end{array} \right) .
\eqno(3.2)
$$
Under the definition (2.3), the neutrino states 
$(\nu_e,\nu_\mu,\nu_\tau)$ are represented by
$$
\left(
\begin{array}{l}
\nu_e \\
\nu_\mu \\
\nu_\tau \\
\end{array} \right)=
\left(\begin{array}{ccc}
\frac{2}{\sqrt6} & \frac{1}{\sqrt3} & 0 \\
-\frac{1}{\sqrt6} & \frac{1}{\sqrt3} & -\frac{1}{\sqrt2} \\
-\frac{1}{\sqrt6} & \frac{1}{\sqrt3} & \frac{1}{\sqrt2} \\
\end{array} \right) 
\left(
\begin{array}{l}
\nu_\eta \\
\nu_\sigma \\
\nu_\pi \\
\end{array} \right) .
\eqno(3.3)
$$
Therefore, it is useful to express the neutrino (Dirac) matrix $M_\nu^D$ 
on the basis $(\nu_\eta,\nu_\sigma,\nu_\pi)$, not on the basis 
$(\nu_\pi,\nu_\eta,\nu_\sigma)$. Here, the neutrino states $\nu_\pi$, 
$\nu_\eta$ and $\nu_\sigma$ are defined by a relation similar to 
Eq.(2.3) with $(\nu_1,\nu_2,\nu_3)=(\nu_e,\nu_\mu,\nu_\tau)$.
When the mass matrix $M_\nu$ becomes diagonal on the basis 
$(\nu_\eta,\nu_\sigma,\nu_\pi)$, the mixing matrix is given by 
the tribimaximal mixing (3.2) exactly.

If we require the S$_3$ symmetry, the Yukawa interaction is generally given 
by the form
$$
H_\nu=y_1\frac{\bar{\ell}_\pi N_\pi+\bar{\ell}_\eta N_\eta
+\bar{\ell}_\sigma N_\sigma}{\sqrt3} \phi^u_\sigma 
+y_2 \left[\frac{\bar{\ell}_\pi N_\eta
+\bar{\ell}_\eta N_\pi}{\sqrt2}\phi^u_\pi
+\frac{\bar{\ell}_\pi N_\pi-\bar{\ell}_\eta N_\eta}{\sqrt2}
\phi^u_\eta \right]
$$
$$
+y_3 \left[\frac{\bar{\ell}_\pi \phi^u_\pi
+\bar{\ell}_\eta \phi^u_\eta}{\sqrt2} N_\sigma+\bar{\ell}_\sigma
\frac{\phi^u_\pi N_\pi+\phi^u_\eta N_\eta}{\sqrt2} \right]
+y_4\frac{\bar{\ell}_\pi N_\pi+\bar{\ell}_\eta N_\eta
-2\bar{\ell}_\sigma N_\sigma}
{\sqrt6}\phi^u_\sigma ,
\eqno(3.4)
$$
where the heavy neutrinos $N_i$ denote $\nu_{Ri}$, and 
$\phi^u= (\bar{\phi}^+_L, \bar{\phi}^0_L)$.
In the present investigation, the S$_3$ symmetry is a necessary
condition, but it is not a sufficient condition.
For the charged lepton sector, we have assumed the
form (3.1), which corresponds to the case with
$y_1=\sqrt{2} y_2 = y_3=y_e/\sqrt{3}$ and $y_4=0$
in the S$_3$ invariant general form (3.4) with
$\bar{\ell}_a N_a \rightarrow \bar{\ell}_a e_{Ra}$
and $\phi^u_a \rightarrow \phi^d_a$ ($a=\pi, \eta, \sigma$).
Therefore, in order to obtain the form (3.1), we must
a further selection rule in addition to the S$_3$ invariance,
for example, a cyclic permutation symmetry, and so on.
For the neutrino sector, in order to reduce the number of
parameters, we assume the following selection rules:

\noindent(i) The VEVs of the up-type Higgs scalars $\phi^u_i$ 
satisfy the relation
$$
\langle \phi^u_\pi \rangle^2+\langle \phi^u_\eta \rangle^2
=\langle \phi^u_\sigma \rangle^2 ,
\eqno(3.5)
$$
as well as Eq.(2.4). (We assume a similar Higgs potential structure for the 
up-type Higgs scalars $\phi^u_i$ as well as $\phi^d_i$.)

\noindent(ii) We have assumed the universality of the Yukawa coupling 
constants for the $(e_1,e_2,e_3)$ basis in the charged lepton sector 
as shown in 
Eq.(3.1). We also assume the universality of the Yukawa coupling constants in 
the neutrino sector  [however, not for the $(\nu_1,\nu_2,\nu_3)$ basis, 
but for the $(\nu_\pi,\nu_\eta,\nu_\sigma)$ basis].

Considering that the observed mixing is almost given by the tribimaximal 
mixing (3.2), i.e. the neutrino Dirac mass matrix $M^D_\nu$ is almost 
diagonal on the 
$(\nu_\eta,\nu_\sigma,\nu_\pi)$ basis, we confine ourselves to 
investigating a 
specific case with $y_1=y_2=y_\nu$ and $y_3=y_4=0$:

$$
H_\nu=y_\nu \frac{\bar{\ell}_\pi N_\pi
+\bar{\ell}_\eta N_\eta+\bar{\ell}_\sigma 
N_\sigma}{\sqrt3}\phi^u_\sigma
+y_\nu \left[\frac{\bar{\ell}_\pi N_\eta
+\bar{\ell}_\eta N_\pi}{\sqrt2}\phi^u_\pi
+\frac{\bar{\ell}_\pi N_\pi
-\bar{\ell}_\eta N_\eta}{\sqrt2}\phi^u_\eta \right] .
\eqno(3.6)
$$
The interaction (3.6) yields the neutrino Dirac mass matrix $M^D_\nu$
$$
M^D_\nu=m^\nu_0 \left(\begin{array}{ccc}
\frac{z^u_\sigma}{\sqrt3}-\frac{z^u_\eta}{\sqrt2} & 0 & 
\frac{z^u_\pi}{\sqrt2}  \\
 0 & \frac{z^u_\sigma}{\sqrt3} & 0 \\
\frac{z^u_\pi}{\sqrt2} & 0 & \frac{z^u_\sigma}{\sqrt3}+\frac{z^u_\eta}{\sqrt2} 
\end{array} \right) ,
\eqno(3.7)
$$
where $m^\nu_0=y_\nu v_u$,  $\langle \phi^u_\pi \rangle=v_u z^u_\pi$,  
$\langle \phi^u_\eta \rangle=v_u z^u_\eta$,  $\langle \phi^u_\sigma \rangle=
v_u z^u_\sigma$, and $(z^u_\pi)^2+(z^u_\eta)^2+(z^u_\sigma)^2=1$. 
The mass 
eigenvalues are given by 
$$
\begin{array}{l}(m^\nu_\eta)'=
\left(\frac{1}{\sqrt3}z^u_\sigma-\frac{1}{\sqrt2}\sqrt{(z^u_\pi)^2
+(z^u_\eta)^2} \right )m^\nu_0, \\
m^\nu_\sigma=\frac{1}{\sqrt3}z^u_\sigma m^\nu_0, \\
(m^\nu_\pi)'=\left(\frac{1}{\sqrt3}z^u_\sigma
+\frac{1}{\sqrt2}\sqrt{(z^u_\pi)^2+(z^u_\eta)^2}
\right)m^\nu_0 . 
\end{array}
\eqno(3.8)
$$
When we use the relation (3.5), i.e.
$$
(z^u_\pi)^2+(z^u_\eta)^2=(z^u_\sigma)^2=\frac{1}{2},
\eqno(3.9)
$$
we obtain
$$
\begin{array}{l}
(m^\nu_\eta)'=(\frac{1}{\sqrt6}-\frac{1}{2})m^\nu_0, \\
m^\nu_\sigma=\frac{1}{\sqrt6}m^\nu_0, \\
(m^\nu_\pi)'=(\frac{1}{\sqrt6}+\frac{1}{2})m^\nu_0, 
\end{array}
\eqno(3.10)
$$
independently of the parameter value of $z^u_\pi/z^u_\eta$
(in other words, independently of the magnitude of the
$\nu_\pi \leftrightarrow \nu_\eta$ mixing). 
The result (3.10) means
$$
\theta_\nu=\frac{\pi}{3},
\eqno(3.11)
$$
from the definition of $\theta_\nu$, Eq.(2.11).

As seen in Fig.~1, although the present interaction form could 
not give the Brannen's relation (2.11), the 
value $\theta_\nu=60^\circ$ is very close to the Brannen's conjecture 
$\theta_\nu=57.73^\circ$. In fact, the result (3.10) predicts
$$
R=\frac{{m_{\nu2}}^2-{m_{\nu1}}^2}{{m_{\nu3}}^2-{m_{\nu2}}^2}
=\frac{4\sqrt6-9}{4\sqrt6+9}=0.04245 .
\eqno(3.12)
$$
Although the value (3.12) is somewhat large comparing with the Brannen's 
prediction (2.14), the prediction(3.11) is, at present, not ruled out
within three sigma.

The value $\langle \phi^u_\pi \rangle \neq 0$ yields the $\nu_\pi$-$\nu_\eta$ 
mixing:
$$
\tan2\theta_{\pi\eta}=\frac{z^u_\pi}{z^u_\eta} .
\eqno(3.13)
$$
If we take $z^u_\pi/z^u_\eta=z^d_\pi/z^d_\eta$, the case yields a large 
deviation from the tribimaximal mixing. Therefore, we cannot choose the same 
values $z^u_i$ as $z^d_i$. We must consider 
$\langle \phi^u_\pi \rangle \simeq 0$ differently from the case of the 
down-type Higgs scalars $\phi^d_i$.

\vspace{3mm}

{\large\bf 4 \ Phenomenological neutrino Yukawa interaction form}

In the previous section, we have required the S$_3$ invariance for
the neutrino Yukawa interaction $H_\nu$, and, in order to obtain
the result (3.11)  near to the Brannen's relation (2.11), we have
found that we need a VEV structure $\langle \phi^u \rangle \simeq 0$
differently from the VEV structure of $\phi^d$.
In the present section, we assume the same structure of 
$\langle \phi^u_i\rangle$ as $\langle \phi^d_i\rangle$ 
(i.e. the same parameter
values of $z_i$).  Alternatively, 
we abandon the S$_3$ invariance of $H_\nu$, although we still use 
the notation $(\pi,\eta,\sigma)$ in the S$_3$ symmetry. 
The purpose of the present section is not to derive the Brannen's 
relation (2.11) from a model 
with some symmetry, but to investigate what Yukawa interaction form is 
required if we take the Brannen's relation (2.11) and the tribimaximal 
mixing (3.2) seriously. Of course, if we introduce several adjustable 
parameters, it is always possible. Therefore, we still adhere to the 
universality of the coupling constants.

We consider only terms which keep the mass matrix diagonal on the 
$(\nu_\eta,\nu_\sigma,\nu_\pi)$ basis:
$$
H_\nu=y_\nu \frac{\bar{\ell}_\pi N_\pi+\bar{\ell}_\eta N_\eta+\bar{\ell}_\sigma 
N_\sigma}{\sqrt3}\phi_\sigma
+y_\nu \frac{\bar{\ell}_\pi N_\pi-\bar{\ell}_\eta N_\eta}{\sqrt2}\phi_x
$$
$$
+y_\nu \frac{\bar{\ell}_\pi N_\pi+\bar{\ell}_\eta N_\eta-2\bar{\ell}_\sigma 
N_\sigma}{\sqrt6}\phi_y .
\eqno(4.1)
$$
Here, we have assumed that the first term in (4.1) is still invariant under 
the S$_3$ symmetry. The three $(\bar{\ell}N)$ terms in (4.1) are linearly 
independent of each other. Similarly, we assume that the scalars $\phi_x$ and 
$\phi_y$ are linearly independent of $\phi_\sigma$, i.e. $\phi_x$ and $\phi_y$ 
are given by linearly independent combinations of $\phi_\pi$ and $\phi_\eta$ .

In order to fix $\phi_x$ and $\phi_y$, 
we assume that
the interaction $H_\nu$ is invariant under the
exchange $\nu_\pi \leftrightarrow \nu_\eta$ 
($\phi_\pi \leftrightarrow \phi_\eta$), i.e. we assume
$$
\phi_x=\frac{\phi_\pi - \phi_\eta}{\sqrt{2}}, \ \ 
\phi_y=\frac{\phi_\pi + \phi_\eta}{\sqrt{2}}.
\eqno(4.2)
$$
Although this symmetry is analogous to the so-called $2\leftrightarrow 3$
flavor symmetry \cite{23sym} in the neutrino mass matrix, the present
$\pi$-$\eta$ symmetry does not mean $\nu_\mu \leftrightarrow \nu_\tau$
symmetry.

As a result, we obtain 
$$
H_\nu = y_\nu \bar{\ell}_\pi N_\pi \left[ \frac{1}{\sqrt{3}} \phi_\sigma
-\sqrt{\frac{2}{3}} \left( -\frac{\sqrt{3}+1}{2\sqrt{2}}\phi_\pi 
+\frac{\sqrt{3}-1}{2\sqrt{2}}\phi_\eta \right) \right]
$$
$$
+y_\nu \bar{\ell}_\eta N_\eta \left[ \frac{1}{\sqrt{3}} \phi_\sigma
-\sqrt{\frac{2}{3}} \left( \frac{\sqrt{3}-1}{2\sqrt{2}}\phi_\pi 
-\frac{\sqrt{3}+1}{2\sqrt{2}}\phi_\eta\right) \right]
$$
$$
+ y_\nu \bar{\ell}_\sigma N_\sigma \left[ \frac{1}{\sqrt{3}} \phi_\sigma
-\sqrt{\frac{2}{3}} \left( \frac{1}{\sqrt{2}}\phi_\pi 
+\frac{1}{\sqrt{2}}\phi_\eta\right) \right] .
\eqno(4.3)
$$
Since we obtain
$$
\langle \phi_\pi \rangle = v z_\pi =
 \frac{v}{\sqrt{2}} \cos\theta_e , \ \ 
\langle \phi_\eta \rangle = v z_\eta =
- \frac{v}{\sqrt{2}} \sin\theta_e , \ \ 
\langle \phi_\sigma \rangle = v z_\sigma =
 \frac{v}{\sqrt{2}}  , 
\eqno(4.4)
$$
from the definitions (2.3) and (2.8), the neutrino
Yukawa interactions (4.3) yields the following 
neutrino Dirac mass matrix $M_\nu^D$ on the $(\nu_\eta,\nu_\sigma,\nu_\eta)$ 
basis:
$$
M_\nu^D ={\rm diag}(m_{\eta\eta}, m_{\sigma\sigma}, 
m_{\pi\pi}) ,
\eqno(4.5)
$$
$$
\begin{array}{l}
m_{\eta\eta} =y_\nu v \left[ \frac{1}{\sqrt{6}}
-\frac{1}{\sqrt{3}} \sin\left( \theta_e +\frac{\pi}{12}
\right) \right] , \\
m_{\sigma\sigma} =y_\nu v \left[ \frac{1}{\sqrt{6}}
+\frac{1}{\sqrt{3}} \sin\left( \theta_e -\frac{\pi}{4}
\right) \right] , \\
m_{\pi\pi} =y_\nu v \left[ \frac{1}{\sqrt{6}}
+\frac{1}{\sqrt{3}} \cos\left( \theta_e -\frac{\pi}{12}
\right) \right] ,
\end{array}
\eqno(4.6)
$$ 
where we have used that $\sin(\pi/12)=(\sqrt{3}-1)/2\sqrt{2}$
and $\cos(\pi/12)=(\sqrt{3}+1)/2\sqrt{2}$.  Comparing the expression (4.6) 
with the expression (2.10),
we obtain the Brannen's empirical relation (2.11), 
$\theta_\nu =\theta_e +\pi/12$, which can give a reasonable prediction (2.13).
Note that the definition of the $(\pi,\eta,\sigma)$ basis
of S$_3$, (2.3), is essential for the derivation of the
Brannen's relation (2.11).
If we took another conventions of the S$_3$ representation,
we would not obtain the Brannen's relation (2.11).

If we regard $m_{\nu 3}$ as $m_{\nu 3}=\sqrt{\Delta m^2_{atm}}$,
from $m_{\nu i} \propto (z_i^\nu)^2$, we obtain
$$
m_{\nu 1}= (3.94^{+0.27}_{-0.43})\times 10^{-4}\, {\rm eV}, \ \ 
m_{\nu 2}= (9.17^{+0.62}_{-0.43})\times 10^{-3}\, {\rm eV}, \ \ 
m_{\nu 3}= (5.22^{+0.35}_{-0.25})\times 10^{-2}\, {\rm eV}, 
\eqno(4.7)
$$
where we have used the best fit value \cite{atm}  
$\Delta m^2_{atm}=(2.72^{+0.38}_{-0.25}) \times 10^{-3}$ eV$^2$.
The predicted values (4.7) lead to
$$
\Delta m^2_{21}=(8.39^{+1.17}_{-0.77}) \times 10^{-5}\ {\rm eV}^2 ,
\eqno(4.8)
$$
which is in good agreement with the observed value
\cite{solar}
$\Delta m^2_{21}=(7.9^{+0.6}_{-0.5}) \times 10^{-5}$ 
eV$^2$.

Of course, since the mass matrix $M_\nu^D$ is diagonal 
in the $(\nu_\eta,\nu_\sigma,\nu_\pi)$ basis, 
the neutrino mixing matrix $U$ is exactly given 
by the tribimaximal mixing (3.2), which gives
$$
\sin^2 2\theta_{23} =1 ,
\eqno(4.9)
$$
$$
\tan^2 \theta_{12} = \frac{1}{2} \ \ \ \ 
(\theta_{12}= 35.26^\circ ),
\eqno(4.10)
$$
$$
|U_{13}|= 0 .
\eqno(4.11)
$$
The prediction (4.10) is also in good agreement with
the observed value \cite{solar}
$\tan^2 \theta_{21}=0.45^{+0.09}_{-0.07}$. We would 
like to emphasize that, in order to obtain the tribimaximal mixing (3.2), 
the magnitudes of the mass eigenvalues $|m_{\eta\eta}|<|m_{\sigma\sigma}|<
|m_{\pi\pi}|$ are essential together with the definition of the S$_3$ basis 
(2.3).

\vspace{3mm}

{\large\bf 5 \ Concluding remarks}

In the present paper, it has been pointed out that if we
take the phenomenological relations (1.2) and (2.11) seriously,
we must consider that the Yukawa interactions in the neutrino sector  
are given by the form (4.1).
The Yukawa interactions (4.1) leads to the tribimaximal mixing
(3.2) and the Brannen's relation (2.11) [so that the neutrino 
mass spectrum (4.7)]. 
Although, in Sec.4, we have put a requirement of 
the $\pi \leftrightarrow \eta$ symmetry for the neutrino 
Yukawa interaction $H_\nu$ instead of the S$_3$ symmetry, 
the requirement is, of course, merely a phenomenological assumption. 
For the charged lepton sector, 
we have assumed the S$_3$ invariant Yukawa interaction, 
while, for the neutrino sector,
we have assumed the S$_3$-breaking interaction (4.1). 
The tribimaximal mixing (3.2) has already been derived,
for example, from an A$_4$ symmetry \cite{Ma06}. 
The interaction form (4.1) will be understood from
an extended finite symmetry.

To the contrary, if we adhere the S$_3$ symmetry, as we discussed in Sec.3, 
we can obtain the result (3.11), which is numerically near to 
the Brannen's relation (2.11). However, we must assume that 
$\langle \phi^u_\pi \rangle \simeq 0$ in the up-type Higgs scalars 
differently from the case in the down-type Higgs scalars. 
The case (3.6) is also interesting.

In the present seesaw model, the 3-family SU(2)$_L$-doublet 
scalars $\phi_i$ cause a flavor-changing neutral current (FCNC)
problem.
Besides, the new heavy fermions $E_{Ri}$ and $E_{Li}$ together
with the scalars $\phi_i$ considerably
affect the evolution of the gauge coupling constants.
If we want to avoid these problems, we can take an
alternative model, a Frogatt-Nielsen-like \cite{Frogatt}
model:
$$
H_{eff}= y_e \bar{\ell}_L H_L^d \left( \frac{\phi}{\Lambda}
\right)^2 e_R 
+y_\nu  \bar{\ell}_L H_L^u \frac{\phi}{\Lambda}
\nu_R  + \bar{\nu}_R M_R \nu_R^* ,
\eqno(5.1)
$$
where $\ell_L =(\nu_L, e_L)$, $H_L^d =(H_d^+, H_d^0)$,
$H_L^u =(H_u^0, H_u^-)$,  $\phi$ is a 3-family
SU(2)$_L$-singlet scalar, and $\Lambda$ is a scale of the
effective theory.
(Here, we have denoted the expression (5.1) symbolically.
For example, the interaction $\bar{\ell} \phi \nu_R$ 
should read the interaction (4.1).)

In the present paper, we have discussed the masses and mixings
only in the lepton sectors.
We think that if there is a beautiful law in the masses and
mixings of the fundamental fermions, we will find it just
in the lepton sectors, because the mass generation
mechanism seems to be simple just in the lepton sectors.
If we define
$$
R_f(\eta_1,\eta_2) = \frac{\frac{2}{3}\left( \eta_1 \sqrt{m_{f1}}
+\eta_2 \sqrt{m_{f2}} + \sqrt{m_{f3}} \right)^2 }{
m_{f1}+m_{f2}+m_{f3}} ,
\eqno(5.2)
$$
where $\eta_i =\pm 1$ ($i=1,2$), the ratio $b/a$ in the
expression (2.5) with $m_{fi} \propto \lambda_i^2$ is
given by
$$
\frac{b}{\sqrt{2} a} = \sqrt{ \frac{2-R}{R} } ,
\eqno(5.3)
$$
where $R$ (not $R_f$) is defined by Eq.(2.13) and $a$ and $b$
are defined by Eq.(2.5) [or Eq.(2.6)].
In the lepton sectors, we have found
$$
R_e(+,+) = R_\nu (-,+) = 1 ,
\eqno(5.4)
$$
so that $(b/\sqrt{2} a)_e = (b/\sqrt{2} a)_\nu =1$.
However, as seen in Table 1, in the quark sectors,
there is no solution of $a/b$ which gives
$R_u (\eta_1^u, \eta_2^u) =R_d(\eta_1^d,\eta_2^d)$.
Therefore, the present idea in the lepton sectors
cannot be applied to the quark sectors straightforwardly.
In the seesaw model (1.5), we have assumed that the heavy
fermion mass matrices $M_F$ in the lepton sectors are
stuructureless, i.e. $M \propto {\bf 1}$.
For quark sectors, we may consider that $M_F$ have
some structure except for the unit matrix ${\bf 1}$,
so that the quark masses will not satisfy the
relation (5.2) with $R_f=1$.
Our goal is to find a unified description of the quark
and lepton masses and mixings.
For this purpose, the present study in the lepton
sectors will provide a promising clue to the unified model.

\vspace{4mm}

\centerline{\large\bf Acknowledgements} 

The author would like to thank C.~Brannen for
communicating his preliminary results prior to
publication.
This work is supported in part by the Grant-in-Aid for
Scientific Research, Ministry of Education, Science and 
Culture, Japan (No.18540284).


\newpage

\vspace{5mm}
\begin{table}
\begin{center}

\caption{ 
Values of $R_f(\eta_1,\eta_2)$ which is defined by 
Eq.~(4.2):
For convenience, the pole mass values \cite{PDG04} for 
the charged lepton masses, 
the values (3.13) for the neutrino masses, and the quark 
mass values for the running mass values at $\mu=m_Z$
\cite{qmass} are used, respectively.
In the Table, the center values have been used as the
input values.
}

\vspace{2mm}
\begin{tabular}{|c|c|c|c|c|}\hline
 Sector & $R(+,+)$ & $R(-,+)$  & $R(+,-)$ & $R(-,-)$ \\ \hline
Charged lepton  & $0.999998$ & $0.946922$ 
& $0.376006$ & $0.34374$ \\
Neutrino & $1.28$  & $1.00$ & $0.251$ & $0.137$ \\
Up-quark & $0.753$ & $0.743$ & $0.590$ & $0.581$ \\
Down-quark & $0.955$ & $0.834$ & $0.481$ & $0.397$ 
 \\ \hline
\end{tabular}

\end{center}

\label{R}

\end{table}
\vspace{2mm}

\newpage
\begin{figure}[h]
\scalebox{0.8}{\includegraphics{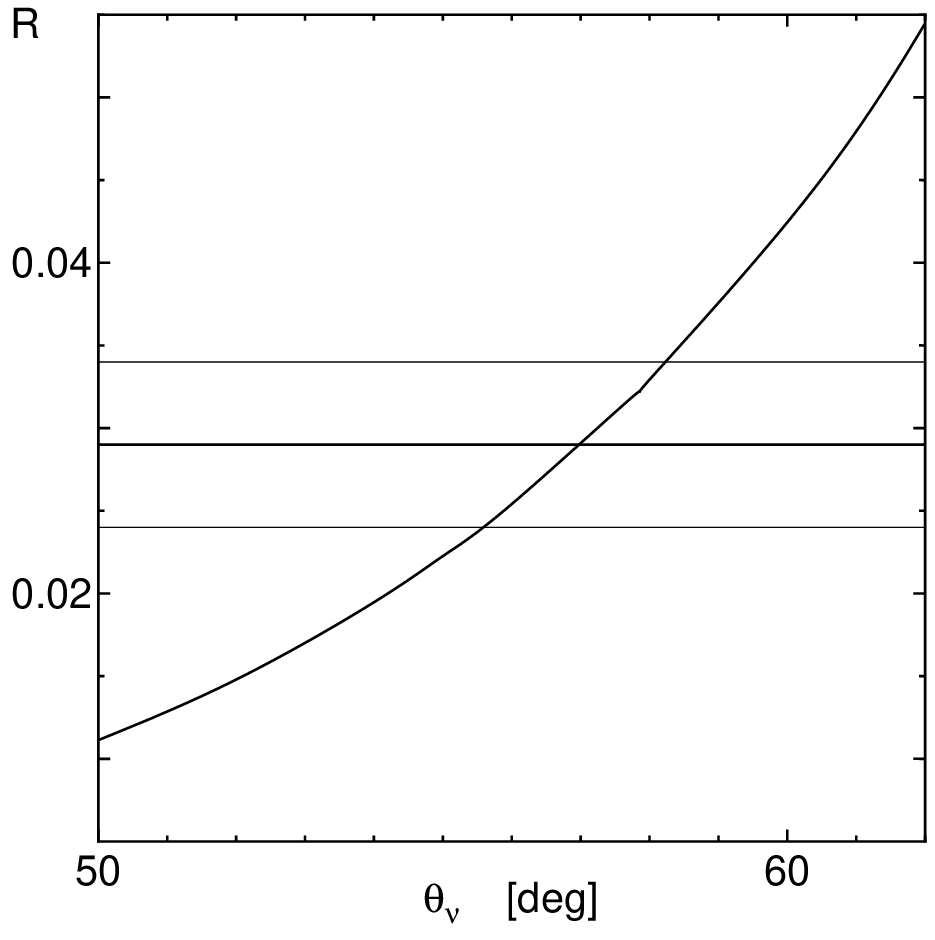}}
\begin{quotation}
{\bf Fig.~1}  $R$ versus $\theta_\nu$ under the relation (1.2).
The horizontal lines denote the observed values
$R =(2.9 \pm 0.5)\times 10^{-2}$.
\end{quotation}

\end{figure}

\end{document}